# Integrated 3D printing of transparency-on-demand glass microstructure

**Zhihan Hong,[1*]† Piaoran Ye,[1*]† Douglas A. Loy,[2,3] Rongguang Liang[1*]**

[1]Wyant College of Optical Sciences, The University of Arizona, 1630 E. University Blvd, Tucson, Arizona 85721, USA

[2]Department of Chemistry&Biochemistry, The University of Arizona, 1306 E. University Blvd, Tucson, Arizona 85721-0041, USA

[3]Department of Materials Science&Engineering, The University of Arizona, 1235 E. James E. Rogers Way, Tucson, Arizona 85721-0012, USA

[*]Corresponding author: Zhihan Hong (zhihanhong@arizona.edu)

Piaoran Ye (piaoranye@arizona.edu)

Rongguang Liang (rliang@optics.arizona.edu)

† These two authors contributed equally to this work.

**ABSTRACT**

Glass is essential in optics and photonics due to its exceptional optical, mechanical, thermal, and chemical properties. Additive manufacturing has emerged as a novel method for fabricating complex glass elements in recent years, yet achieving locally controlled transparency in glass micro-objects remains a significant challenge. We present an innovative method, termed Transparency-on-Demand Glass Additive Manufacturing (TGAM), to control the transparency of 3D printed glass elements using polymeric silsesquioxane (PSQ) and two-photon polymerization (TPP). By precisely manipulating key parameters such as laser power, scanning speed, part thickness, and pyrolysis heating rate, we achieve the desired transparency levels. Our study reveals that monomer conversion during printing, structure thickness, and pyrolysis heating strategy significantly influence PSQ oxidation, resulting in varying transparency in the final glass product. This method enables the creation of high-precision, variable-transparency glass micro-components, providing a scalable and efficient solution for producing complex glass structures with tailored optical transparency. Our technique paves the way for integrated manufacturing of controllable-transparency glass micro-structures, unlocking new possibilities for advanced optical and photonic applications.

**Introduction**

Glass is a ubiquitous material in modern society, renowned for its exceptional optical, mechanical, thermal, and chemical properties. Its versatility and durability have led to its widespread use in numerous applications, including optics and photonics, architecture, automotive, consumer electronics, medical, energy, household items, aerospace, industrial, and environmental applications due to its exceptional optical, mechanical, thermal, and chemical properties (*1-10*). However, compared to metals and polymers, Glass elements are more challenging to fabricate, particularly when creating complex and precise geometries (*11*). Recently, additive manufacturing, or 3D printing, has emerged as a revolutionary method for fabricating glass elements. Despite its advantages, current 3D printing techniques for glass elements suffer from limited control over their transparency(*12-23*). This limitation restricts the applications requiring high transmission and high contrast, such as telecommunications(*24, 25*), remote sensing(*26, 27*), integrated photonics(*28, 29*), nanosatellites(*30*), biomedical optical imaging(*31-36*), optical sensors(*37, 38*), biology sensors(*39*), UV absorbers(*40, 41*), thermal isolators(*42-44*), and neutral-density filters(*45, 46*) and consumer electrical devices(*47*). Tailoring transparency in glass will not only enhance the system performance, but also broaden its application range. For instance, a high-performance imaging system uses aperture and stop to minimize stray light for high contrast imaging, the system can be much more compact if the aperture and stop can be built directly into the optical elements during the fabrication process.

The transparency of glass can be controlled by adding transition metal ions(*48*). The transparency of the glass is highly dependent on the number of additives and the thickness of the structure. The limitation of this method is that it lacks local transparency control, especially for nano- and micro-scale features. Furthermore, the addition of inorganic metal ions complicates the direct laser writing (DLW) process by decreasing light penetration. The concentration of metal ions in glass 3D printing materials is also typically limited due to the poor solubility of metal salts in printing resins (*49*). To date, 3D printing of inorganic glass/ceramic micro-objects with controlled transparency has not been reported.

Controlling glass purity without additives is one potential approach to achieve tailored transparency in high-precision 3D printed glass. While powder-based printing materials using high-purity silica nanopowder are difficult to tune due to SiO2's stable properties (*13*), silsesquioxane-based materials offer more flexibility owing to their carbon-rich nature. The process to fabricate glass from silsesquioxane-based materials typically involves photo-polymerization printing, washing, and high-temperature treatment. During printing, carbon bonds form through cross-linking to create 3D shapes (*50-53*). Subsequent pyrolysis in air usually cleaves and oxidizes hydrocarbons bonded to Si atoms, leaving only Si and O. However, incomplete oxidation results in residual carbon species, which can cause a loss in transparency(*11*). By controlling the content of unoxidized carbon species during pyrolysis, it is possible to produce glass or ceramics with controlled transparency and colored appearances.

In this work, we present a method utilizing two-photon polymerization (TPP) and polymeric silsesquioxane (PSQ) to 3D print inorganic glass/ceramic micro-objects with controlled transparency. The transparency of these objects can be regulated through multiple mechanisms: adjusting the sample thickness, optimizing pyrolysis conditions, tuning the laser power and scanning speed. This process offers the flexibility to create single glass/ceramic objects with varying transparency regions, a capability that has been challenging and previously unachievable for glass micro-objects. This approach expands the capabilities of 3D printing technology for producing complex glass and ceramic structures with tailored optical properties. We named this technology Transparency-on-Demand Glass Additive Manufacturing (TGAM).

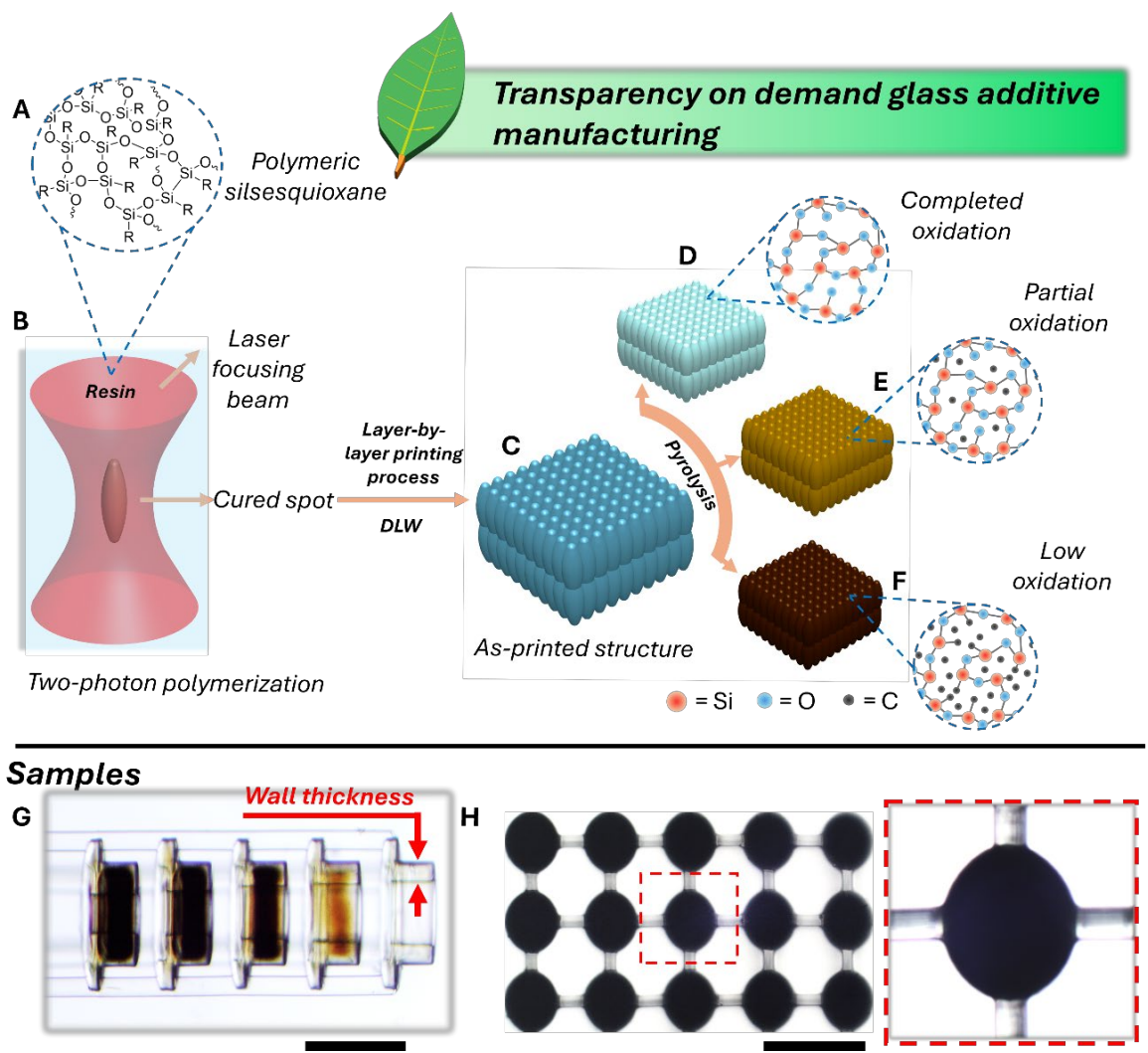

Figure 1. (**A**) Structure of polymeric silsesquioxane. (**B**) TPP (Two-photon polymerization) curing process. (**C**) 3D printed structure. (**D**)-(**F**) The transparency performances under different degrees of oxidation. (**G**) Multi-transparency ring tower. (**H**) Ball-and-stick structure. (scale bar: 100 μm)

**Results and Discussion**

The material is based on PSQ (*14, 54*) (Fig. 1A). TPP (two-photon polymerization) (Fig. 1B) can be utilized to achieve sub-wavelength spatial resolution control for complex micro geometrics. After printing and washing, the printed objects are pyrolyzed in air at 650 °C to reach inorganic status. The transparency of the final glass is determined by the ability of oxygen to penetrate and react with the organic components embedded in the cured PSQ.

Before we print glass objects with various transparencies, we explore whether the PSQ can reach different transparencies after pyrolysis in air by controlling the oxidation degree. Two batches of free stand thin films were prepared: thin films pyrolyzed in air and thin films pyrolyzed in a high-vacuum environment, which limits the oxidation of hydrocarbons in the cured PSQ material. The samples treated in air were colorless and transparent, while those treated under vacuum were black and exhibited very low transparency to visible light. Solid state NMR was used to validate that the dark or brown tint is primarily caused by the unoxidized carbon composition after pyrolysis. The $^{29}$Si NMR spectrum revealed that the primary structures of Si in the transparent samples were Q3 and Q4 types, with a minor presence of Q2 type (Figure S1A, from -90 ppm to -120 ppm). Notably, no T-type peaks were observed, indicating that almost all carbons were cleaved from the Si atoms. Additionally, the $^{13}$C NMR spectrum showed no aromatic species remaining in the structure, suggesting a high level of oxidation was achieved (Figure S2A). Conversely, the samples pyrolyzed under vacuum displayed T-type peaks from -45 ppm to -80 ppm, indicating that some C atoms were still connected to Si atoms, likely due to incomplete oxidation (Figure S1B). Furthermore, the broad peak (~105 ppm – 150 ppm) in the $^{13}$C NMR spectrum confirmed the presence of aromatic species in the black samples, which is common in pyrolysis when oxidation is incomplete (Figure S2B).

We then investigated whether the oxidation degree (and thus transparency) of the pyrolyzed printed parts could be adjusted before they are pyrolyzed in air. The most straightforward method to control oxidation is by manipulating sample thickness, as thicker samples impede oxygen penetration and complete oxidation. Figures 1(G)&(H) illustrate this effect. Figure 1(G) showcases a ring tower with varying wall thicknesses (from left to right), clearly demonstrating variations in transparency. Figure 1(H) presents a ball-and-stick model where balls with a diameter of approximately 50 μm appear entirely black, while sticks with a thickness of 10 μm are transparent. These results indicate that even when pyrolyzed in air, printed parts can achieve low transparency because the cured PSQ is sufficiently dense to prevent oxygen permeation in thicker sections.

With that, we further explored how the transparency of the final glass can be finely regulated by tuning various fabrication and process conditions. To investigate this, we designed and printed a testing cylinder plate structure (Figs. 2A&B) with supporting rings and pillars to create distances between the testing area and the quartz substrate, thereby avoiding interference that could affect the final transparency. The testing area consisted of a cylinder plate with a diameter of 125 μm, produced in three different thicknesses (16 μm, 17 μm, and 18 μm). For each thickness, we applied various heating rates during pyrolysis, as shown in Figures 2C&D. We also discovered that the final transparency could be tuned by adjusting the laser power and scanning speed during printing. Consequently, for each thickness and heating rate combination, we further applied six different laser pulse energies and scanning speeds (Figure 2D), resulting in 324 unique transparent levels across all parameter combinations. This comprehensive approach allowed for a thorough investigation into transparency control.

As expected, thicker cylinder plates were more likely to yield glass with brown or dark coloration, given the same printing power, scanning speed, and pyrolysis heating rate. Notably, transparency was extremely sensitive to thickness differences, allowing for efficient adjustment within a range of 1 to 2 micrometers. The heating rate during pyrolysis also significantly influences the degree of hydrocarbon oxidation. In our pyrolysis process, all samples were initially heated to 300 °C at a consistent rate of 6.875 °C/min. Subsequently, different groups were heated to 650 °C at rates ranging from 5.470 °C/min to 4.118 °C/min, respectively (Fig. 2C). Figure 2D demonstrates that samples heated at higher rates were more prone to turning brown and black. While directly monitoring pyrolysis kinetics under varying heating rates is challenging, we propose that higher heating rates cause the object's shell to pyrolyze more rapidly compared to the oxygen permeation rate. Once a condensed silica shell forms, it inhibits further oxygen permeation, halting oxidation within the object and resulting in a colored appearance. Conversely, slower heating rates promote a more homogeneous pyrolysis process throughout the object, facilitating complete oxidation.

Furthermore, one of the most intriguing and powerful processes we discovered for manipulating transparency is adjusting the laser power and scanning speed during printing. Although these parameters shouldn't alter the chemical structure, as they all lead to photo-induced free radical polymerization, they significantly affect the monomer conversion (crosslinking degree) of the final printed objects. This influences oxidation levels and transparency in two ways: 1) Higher power and lower scanning speed result in higher monomer conversion, creating denser material as less monomer is washed away after uncured material removal (*55*), which makes oxygen permeation more difficult. 2) A higher crosslinking degree of acrylates generally leads to a higher degradation temperature, making complete oxidation more difficult (*56*).

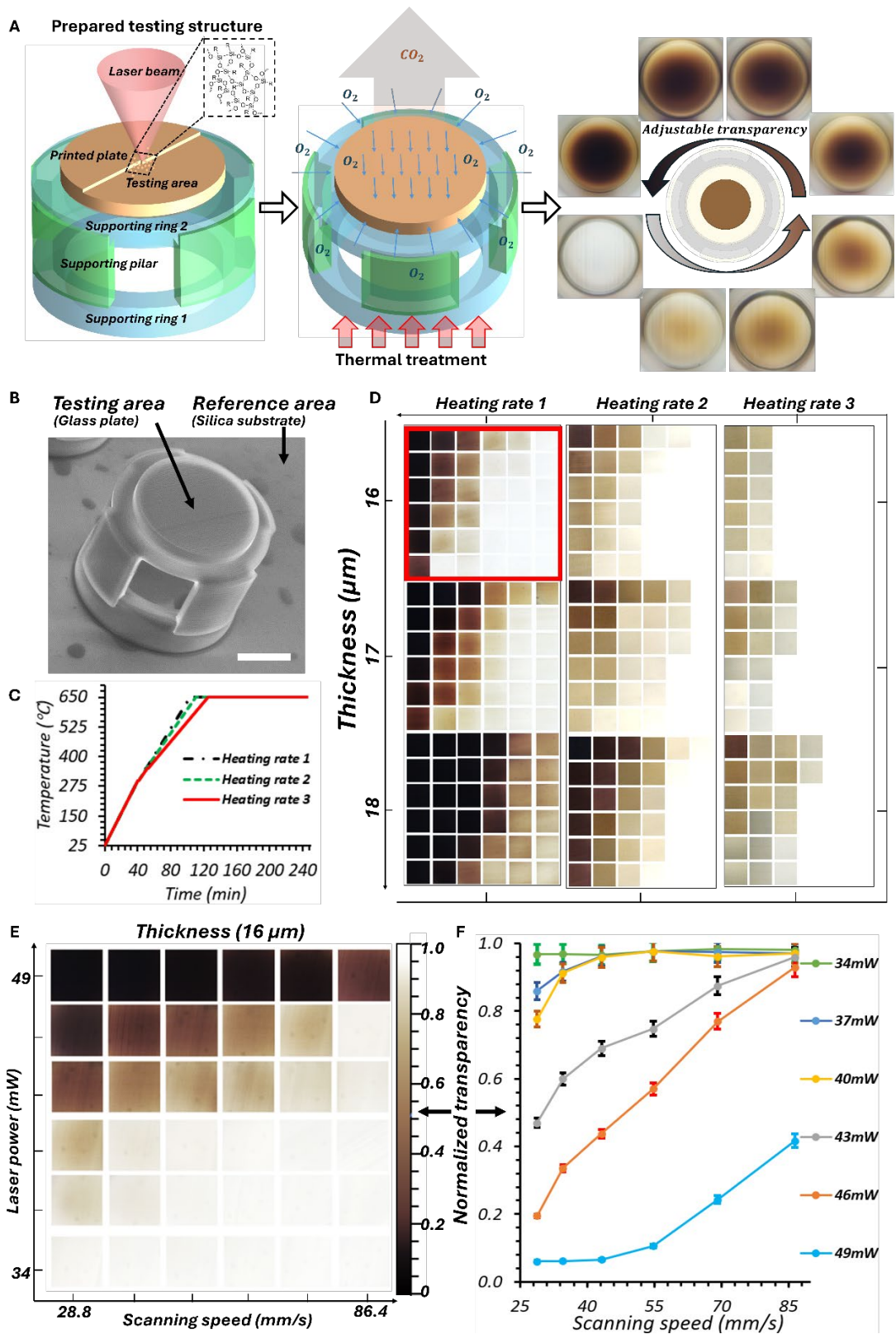

Figure 2. (**A**) The designed cylinder plate, the thermal treatment process and the final glass with different transparency. (**B**) The SEM image of the as-printed structure (scale bar: 50 μm). (**C**) The three different heating

strategies in (**D**). (**D**) The transparency information is obtained with various laser pulse energy, scanning speed, plate thickness, and heating rate. (**E**) As printed 16 μm thickness cylinder flat plate after pyrolyzed at 'Heating rate 1'. (**F**) The normalized transparency in (**E**).

To demonstrate the feasibility of manipulating transparency by changing printing parameters, we achieved local control over the transparency of pyrolyzed glass by varying the laser power and scanning speed during printing, as shown in Figs. 2D&E. All cylinder plates are visually identical right after printing. Fig. 2E illustrates a printed glass cylinder plates with an as-printed structure thickness of 16 μm and pyrolyzed at "Heating rate 1" (Fig. 2C). To normalize transparency measurements, we used the captured light intensity of a structure-free area on the same quartz substrate as a reference (Fig. 2B). The resulting transparency is depicted in Fig. 2E&F, showing that transparency decreases with increased laser pulse energy and decreased laser scanning speed. For the cylinder plate with a thickness of 16 μm before pyrolysis, the lowest transmitted intensity was approximately 4% (laser power of 49 mW, scanning speed of 28.8 mm/s). This strategy allows us to pre-program the transparency during the printing in different regions of one object even with the same thickness. It's important to note that higher laser power and slower scanning speed increase the size of polymer features, potentially leading to greater thickness that can also affect transparency. To ensure consistent thickness across test plates, we adjusted the z-axis travel distance for compensation, allowing independent control of both thickness and monomer conversion.

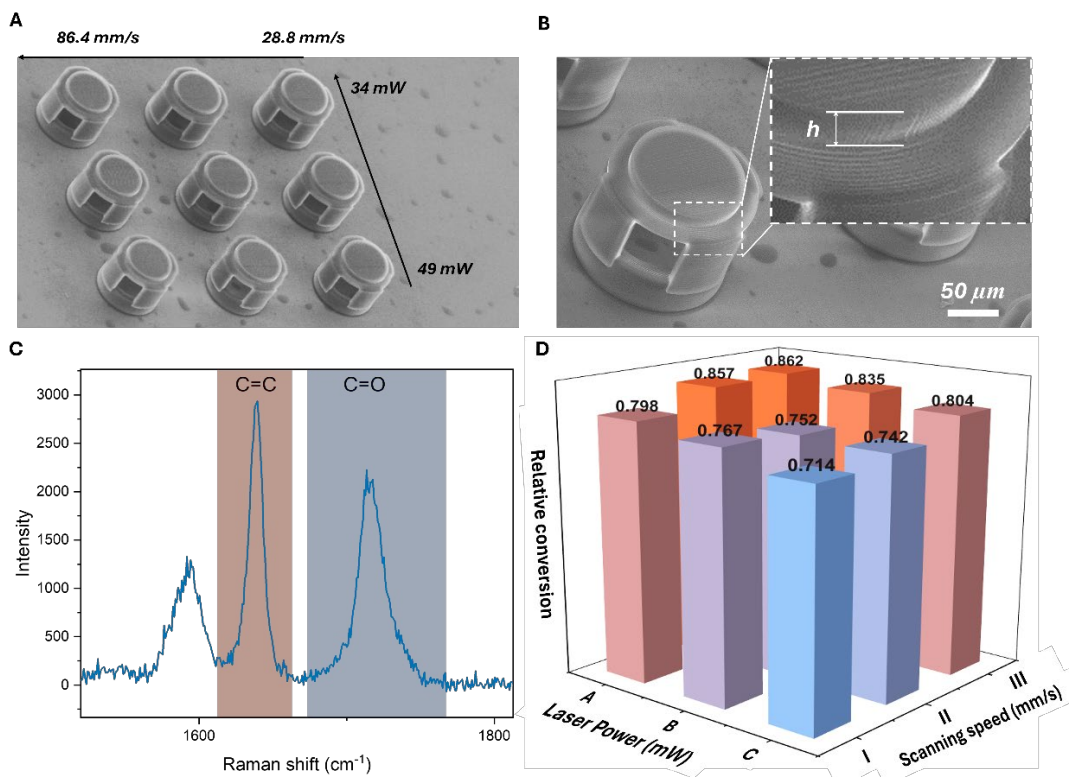

Figure 3. (**A**)&(**B**) SEM of testing cylinder plates printed with different scanning speeds and laser pulse energies. (**C**) Micro-Raman spectrum of one test plates in (A) after printing and washing process. (**D**) The calculated relative

unreacted monomer conversion. Group A, B, and C refer to laser power of 34 mW, 43 mW, and 49 mW, respectively. Group I, II, and III refer to scanning speeds of 28.8 mm/s, 43.2 mm/s, and 86.4 mm/s, respectively.

Raman spectroscopy was employed to demonstrate monomer conversion differences between cylinder plates printed with varying parameters. Nine plates of identical thickness (16 μm, confirmed by white light interferometry, shown in Fig. S3) were printed using different parameters, as shown in Figure 3A. Raman microspectroscopy quantitatively measured the vibration of unreacted C=C bonds (1639 cm$^{-1}$) and the constant C=O bonds (cm$^{-1}$), as illustrated in Figure 3D. We used the area ratio between C=C and C=O to represent the relative unreacted monomer (the lower the value, the higher the monomer conversion), as the carbonyl group remains unchanged during polymerization, except for a negligible amount from the initiator (<1%).Figure 3E demonstrates that as laser power increases and scanning speed decreases, the relative unreacted monomer decreases, reaching a mininum value of 0.714 during the test. This corresponds to the highest power of 49 mW and the slowest speed of 28.8 mm/s. Notably, cylinder plate B-II shows a slightly higher monomer conversion than B-I, likely due to laser power fluctuations during the experiment, which proved difficult to completely eliminate with our experimental setup. Post-pyrolysis observations revealed that cylinder plate B-II exhibited a darker appearance compared to B-I, aligning with the relative conversion results (Figure S4). This demonstrates that the final transparency is directly linked to the monomer conversion.

It is important to note that variations in oxidation degree not only affect transparency but also lead to different shrinkage rates due to varying amounts of carbon removed during pyrolysis. For the samples studied in Figure 2D, the shrinkage ranges from 29% to 32.4%, corresponding to transparency levels from approximately 4% to 95%.

The knowledge established above enables us to flexibly adjust the transparency of printed micro-objects, creating either identical shapes with varying transparencies or complex structures with multiple regions of different transparencies. For example, we printed black and white swans in an embrace, demonstrating uniform transparency from head to body and distinct white flight feathers on the black swan's wings (Fig. 4A). We also printed binary bars, a checkerboard pattern, and a black aperture on a clear plate, showcasing pre-designed transparency variations through tuning feature thickness or conversion degree (Fig. 4B).

Our printed structures performed well without cracks, although slight shrinkage differences led to minor deformations. Figures 4C&D illustrate printed tubes with external diameters (ED) of 220 μm and 155 μm, respectively, both with a wall thickness (WT) of 17 μm, subjected to "Heating rate 1" thermal treatment (see the supplementary materials for details). Beyond the color (transparency) differences observed under microscopic imaging, SEM analysis revealed ED differences of the regions with higher and lower transparency. The black and yellow tube (Fig. 4C) showed an ED difference of approximately 1.2% ± 0.1%, while the black and white tube (Fig. 4D) exhibited an ED difference of about 3.1% ± 0.2%. These differences, primarily caused by varying shrinkage, can potentially be compensated for in the original design.

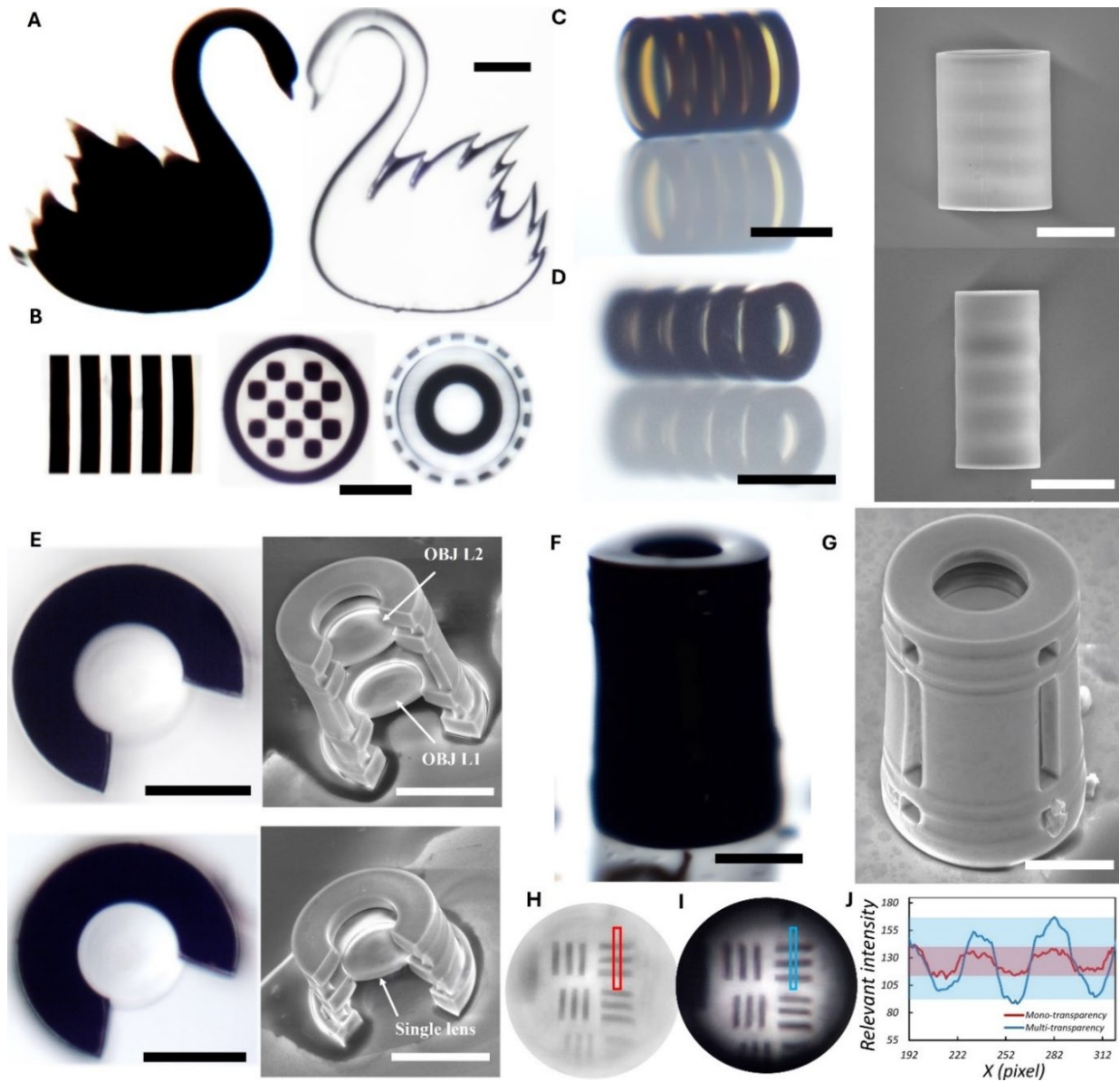

Figure 4. (**A**) Black & white glass swan images captured by transmission microscope (scale bar: 25 μm). (**B**) Glass binary bars, checkboard, and circle ring structure images captured by transmission microscope (scale bar: 100 μm). (**C**)&(**D**) Black and yellow ring pattern glass tube and black and white glass tube (scale bar: 100 μm). (**E**) 3/4 doublet and singlet imaging glass optical system with integrated black aperture (scale bar: 100 μm). (**F**) Integrated additive manufactured glass objective including tube, mount, stop and apertures (scale bar: 100 μm). (**G**) The SEM image of the integrated objective lens in (**F**). (**H**) Imaging performance of totally transparent glass objective lens without aperture structure. (**I**) Imaging performance of integrated glass objective lens shown in (**F**). (**J**) The red and blue rectangles in the images mark the area used for the contrast comparison.

Furthermore, we addressed the issue of stray light suppression in micro-optical systems (Figs. 4E-4J). Stray light significantly reduces contrast in imaging systems, posing challenges in fields such as astronomy, medical testing, and electronic equipment. Current solutions involve adding special components like light shields and apertures, but these require additional assembly and lack efficiency, especially for glass micro-optics. Using our transparency-controlled glass 3D printing technique, we can directly fabricate integrated optical components containing lenses, apertures, stop rings, and mounts. Figure 4E shows a 3/4 doublet and singlet imaging glass optical systems (Diameter: 200 μm) with visible light images from top-view and SEM images from side-view, demonstrating clear glass lenses and dark opaque fixtures. Figures 4F & G display the complete optical system, featuring a thin gap on the side to facilitate washing out of uncured liquid material after printing. A profile comparison of two intensity cuts through a line pair of the USAF test target for the singlet lens, with and without a black aperture, revealed an enhancement in the contrast (CM = (Imax − Imin)/(Imax + Imin)) by more than a factor of 3.5. This demonstrates the effectiveness of our method in achieving desired transparencies in 3D-printed glass components, enhancing their application in various fields requiring specific optical properties.

In summary, by controlling the oxidation degree of printed PSQ during printing and pyrolysis, we have developed a transparency-on-demand glass 3D printing technique. This method enables the direct manufacturing of multi-transparency micro 3D architectures without the need for any additives and other processes. The TGAM technique employs laser-induced conversion adjustments and thickness variations during the printing process, as well as the heating rate during the pyrolysis, to manage the transparency distribution of the glass, enhancing the functionality and versatility of the printed structures. We believe it broadens the scope of light-material interaction dimensions, paving the way for more complex and diverse applications.

**Author contributions**

Z.H. and P.Y. contributed equally to this work. Z.H., P.Y., D.A.L., and R.L. conceived the idea and designed the study. P.Y. and D.A.L. designed the material. P.Y. and Z.H. prepared and characterized the materials. Z.H. designed the structures, performed printing experiments, and conducted the imaging experiments. Z.H., P.Y., D.A.L., and R.L. analyzed and interpreted the result, and wrote the manuscript.

**Acknowledgment**

This work was supported by National Cancer Institute R21CA268190.